\documentclass[11pt]{article}
\usepackage{aaspp4}
\tighten
 
\def\gtsima{$\, \buildrel > \over \sim \,$}
\def\ltsima{$\, \buildrel < \over \sim \,$}
\def\simgt{\lower.5ex\hbox{\gtsima}}
\def\simlt{\lower.5ex\hbox{\ltsima}}

\def\HST{{\it HST}}
\def\IUE{{\it IUE}}        	
\def\pmb#1{\setbox0=\hbox{#1}
  \kern-.02em\copy0\kern-\wd0
  \kern.01em\copy0\kern-\wd0
  \kern.01em\copy0\kern-\wd0
  \kern.01em\copy0\kern-\wd0
  \kern.01em\copy0\kern-\wd0
  \kern-.02em\raise.01em\box0 }
\def\kms{{\>\rm km\>s^{-1}}}
\def\Teff{T_{\rm eff}}

\begin{document}

\title{\leavevmode\pmb{\it Hubble Space Telescope\/} 
Observations of the Planetary Nebula K\,648 \\
in the Globular Cluster M15\footnote{Based on observations with 
the NASA/ESA {\it Hubble Space Telescope}, obtained at the Space
Telescope Science Institute, which is operated by AURA, Inc., under
NASA contract NAS5-26555.} }
 
\bigskip

\author{David R. Alves, Howard E. Bond, and Mario Livio}

\affil{Space Telescope Science Institute, 
3700 San Martin Dr., Baltimore, MD 21218 \\
Emails: {\tt alves, bond, mlivio@stsci.edu} }

\clearpage

\begin{abstract}

We have obtained observations of the planetary nebula K\,648 in the Galactic
globular cluster M15 with the {\it Hubble Space Telescope's\/} WFPC2 camera, 
covering an interval of 7~days. The frames
provide
both time-sampled broad-band photometry of the central star and 
high-resolution images of the nebula in the light of H$\alpha$,
[\ion{O}{3}], and [\ion{N}{2}].  

In the deep narrow-band
images, K\,648 is 
a fairly typical double-shelled
elliptical, but with a bright arc at one end of the major
axis that is especially prominent in [\ion{N}{2}]; this feature is probably a
collection of FLIERs (fast low-ionization emission regions).  The nebula is
surrounded by a faint, smooth elliptical halo, which appears undisturbed by
any interaction with the interstellar medium.

Adopting $\Teff = 40,000 \pm 3,000$~K
based upon published spectral-line analyses, 
and
employing our new broad-band optical flux data along with
the known cluster distance,
we  find $\log L/L_{\odot} = 3.78 \pm 0.08$
for the K\,648 central star.
Theoretical 
post-asymptotic-giant-branch evolutionary tracks 
imply a mass of
$0.60 \pm 0.02 \, M_{\odot}$ for this luminosity,
which is significantly higher than the mean mass 
of white dwarfs in globular clusters and the halo
field ($0.50 \pm 0.02\, M_{\odot}$).
The K\,648 central star exhibits no significant 
photometric variability in our data;
thus we find no direct evidence of a 
close binary companion.
We suggest that the progenitor of K\,648 
experienced mass augmentation in
a close-binary merger, allowing it to evolve
to a remnant of higher mass than those of the single stars 
in the cluster.

\end{abstract}


\clearpage

\section{Introduction}

A small number of planetary nebulae (PNe) are associated with the Galactic
halo, on the basis of 
their kinematics, Galactic locations, chemical compositions, and/or
memberships in globular clusters.  These halo PNe (called Type~IV by Peimbert 
1978) are of particular interest
because they arise from the low-mass and low-metallicity extremes among PN
progenitor stars, and are thus important objects for testing our understanding
of PN formation. 

Eight halo PNe are known in the field
(see Howard, Henry, \& McCartney 1997 and references therein).
Four additional
PNe are known in Galactic globular clusters (GCs), including the subject of
this paper and the prototype of the class, K\,648 in the cluster M15
(NGC~7078). First cataloged as a star by K\"ustner (1921), K~648 was shown to
be a PN by Pease (1928).  During the ensuing six decades K~648 remained 
unique, but
recently PNe have also been found in the GCs M22 (Gillett
et~al.\ 1989) and Pal~6 and NGC~6441 (Jacoby et~al.\ 1997). 

It is currently believed that PNe may form through two mechanisms: envelope
ejection from single red giants, 
or common-envelope ejection from close binaries. The relative importance of
these mechanisms, particularly as a function of progenitor mass, is not well
known.

However, observational evidence is mounting that
binary-star processes are a key element in the origin of halo PNe.  
As noted above, the Jacoby et~al.\ (1997) survey of 133 GCs for PNe brought
the total number known to four. These authors pointed out 
that this is 
3$\sigma$ lower than the $\sim$16 PNe
expected on the basis of the total luminosity of the 
Galactic GC system, a PN lifetime of 
$\sim$2--$3\times10^4$~yr, and the assumption
that every star produces a visible PN\null.
In order to explain the small number actually observed,
Jacoby et~al.\ (1997) suggested that 
the last assumption may be incorrect in GCs.  
In fact, 
the post-asymptotic-giant-branch (post-AGB) stellar evolutionary
timescales of individual stars in GCs
may be so long that
the ejected nebular material has time to disperse before the central star 
becomes hot enough to ionize it. (This has been called ``lazy'' evolution by, 
e.g., 
Renzini 1981.)

The notion of lazy post-AGB evolution for GC and
halo stars is supported by several
lines of evidence.  First,
recent {\it Hubble Space Telescope\/} (\HST\/)
detections of white dwarfs in nearby GCs
(Renzini et al.\ 1996; Cool, Piotto, \& King 1996;
Richer et~al.\ 1997) indicate that the remnant
masses are very low
($0.50\pm0.02\,M_\odot$; see \S7).
Post-AGB stars of masses $\simlt0.55M_\odot$
have theoretical evolutionary 
timescales that are much longer than the dissipation timescale for PNe
(Sch\"onberner 1983; Vassiliadis \& Wood 1994).
In addition, Bond \& Fullton (1997, 1999) have used
star counts of post-AGB stars in the M31 halo to
estimate a lifetime of
$\sim$25,000 years for the portion of the post-AGB evolution
from effective temperatures of 5,000 to 10,000~K\null.
Since this is already comparable to the nebular dispersion time,
it is unlikely that these halo stars will ionize visible PNe.

The problem then shifts from ``Why are there so few PNe in globular 
clusters?''\ to ``Why are there {\it any\/} PNe at all?''.
One possibility is that there might be enough scatter in the 
initial-mass\slash final-mass relation to produce occasional remnants that 
are massive enough to produce PNe. 
Jacoby et al.~(1997), however,
proposed that the known GC PNe were formed
through close-binary interactions, although they did not propose any specific 
mechanisms.
As mentioned above,
the ejection of PNe by close-binary stars is 
generally attributed to a common-envelope (CE)
interaction (Pacz\'{y}nski 1976; 
Bond \& Livio 1990; see Iben \& Livio 1993 for a review).
In this scenario, one component of the binary is able to
grow to red-giant dimensions,
and only then swallows its companion into an envelope surrounding both stars.
The orbit spirals down as orbital energy is transferred to the envelope,
which can ultimately be spun up and ejected.  

Several close-binary evolutionary channels may lead to formation of
PNe in GCs, and these PNe may have either single 
central stars or central stars that are still binaries.
In the standard terminology of binary interactions,
Case~B, Case~C, and probably also
Case~A interactions, are all of interest.  

In a Case~B interaction,
the CE event occurs while the primary 
is on the first-ascent red giant branch (RGB), and yields a 
helium white dwarf with a main-sequence secondary.  
The white-dwarf remnant of a Case~B ejection will probably
evolve very slowly toward high temperatures, and thus will not
form a visible 
PN (another example of lazy evolution).  However, in extreme
cases of mass loss during the CE event, a PN with a low-luminosity
close-binary central star may be formed (Iben \& Tutukov 1993).
In a Case~C interaction, the
CE event does not occur until the primary has reached the 
AGB, and yields a carbon-oxygen
white dwarf with a main-sequence secondary.  
If the primary's initial mass was sufficiently high, the CO white dwarf
from a Case~C ejection will evolve rapidly towards high temperatures,
and form a PN with a close-binary central star 
(Iben \& Tutukov 1993).  
Another possible outcome of either a Case~B or C interaction
is that the
CE event results in a stellar merger.
The product of two merged GC stars
may be a rapidly rotating red giant
(Livio \& Soker 1988), which may
eventually become a PN by evolving as a higher-mass single star
(Yungelson, Tutukov, \& Livio 1993).
Han, Podsiadlowski, \& Eggleton (1995) 
have used population synthesis simulations to 
predict
the probabilities for these scenarios in a low-mass,
low-metallicity field population consisting initially
of 50\% binaries:
$\sim$10-15\% Case~B, $\sim$1-4\% Case~C, and $\sim$7\% mergers (the majority 
of the binary population having an initial separation too large for any CE 
interaction to occur).  

There is yet another way for binary-star interactions to give rise
to halo PNe, especially in a GC\null. That would be through a
Case~A interaction, which occurs while both components are on
or near the main sequence (and thus a CE event does not occur).  
Such an interaction could involve a binary that 
goes into contact (i.e., a W~Ursae Majoris system) and coalesces (e.g.,
Webbink 1976).  The coalesced star would appear
first as a blue straggler lying near the main sequence but above the cluster's 
turnoff,
and it would evolve subsequently as a single star of higher mass and shorter
evolutionary timescale than stars at the main-sequence turnoff.  When such a
star ejects its envelope and leaves the AGB, it would produce a relatively
massive post-AGB remnant which could form a PN.  Many
blue stragglers, both in GCs (e.g., Kaluzny et~al.\ 1997 and
references therein) and in the field (e.g., Bond \& MacConnell 1971)
are known binaries, which have so far avoided merging.
This may suggest that mass transfer without a merger is what
augments the mass of the blue straggler.
It is also possible that binary blue stragglers are the
products of a triple-star interaction
(Iben \& Tutukov 1999) or a binary/binary interaction, with ejection of one
star from the system, or with one pair merging and the third star 
left as a wider companion.  
In any case, it is now well-known that most GCs contain
blue stragglers, presumably formed through collisionally induced
mergers that occur in the cluster
core, so the above scenario for producing PNe in GCs is plausible
(cf.\ Livio 1993; 
Stryker 1993; Sills \& Bailyn 1999; and references therein). 
However, given the total numbers of blue 
stragglers found in GCs, 
it is not clear that they are sufficient to produce 
the observed number of GC PNe.

In summary, we have outlined a variety of close-binary evolutionary channels
which could produce PNe in halo populations.
In some of these scenarios, the central star of K\,648 would
still be a close binary; in others the star would be a merged remnant.
In the field,
$\sim$10--15\% of PNe have photometrically
detectable close-binary central stars 
(Bond \& Livio 1990; Bond 2000).
In most of these cases, the photometric variability
is due to heating of a main-sequence companion
(actual eclipses thus not being necessary for detection
of binarity).   Moreover, the short periods of the binary central
stars detected in this manner require the progenitor systems
to have evolved through a CE phase. Thus, if we could show that the nucleus of 
K\,648 is such a close binary, this PN would become an important example of CE 
evolution in the halo population.
On this basis,
we undertook a photometric monitoring campaign on K\,648
with the Wide Field Planetary Camera~2 (WFPC2) 
camera onboard the 
\HST\ in order to search for a close binary central star.

This paper is organized as follows.
In \S2, we review relevant observational
studies of K\,648 and its host cluster, M15.
We summarize our new \HST\/ observations in \S3 and 
in \S4 we describe our photometry of the K\,648 central star.
We derive the fundamental parameters of the
central star in \S5, and
discuss the morphology of the nebula in \S6.
In \S7, we discuss the evolutionary status of
the K\,648 central star.   
We conclude in \S8.

\section{K\"{u}stner\,648 in Messier 15}

K\,648 lies in a crowded field near the center of M15.  In the PN literature, 
it is also cataloged as Ps~1 and PN~G065.0$-$27.3 (Acker et~al.\ 1992).
Radial-velocity measurements confirm its membership in M15 (Joy 1949).
In this paper, we will refer to the nebula as K\,648, and to
its central star explicitly.

O'Dell, Peimbert, \& Kinman (1964) 
presented the first
detailed spectrophotometric study of K\,648.  
Subsequent spectrophotometric studies
have been reported by Miller (1969), Peimbert (1973), 
Hawley \& Miller (1978), Torres-Peimbert \& Peimbert (1979), Barker (1980),
and Adams et al.~(1984).  
Johnson, Balick, \& Thompson (1979) resolved K\,648 with the Very Large Array
radio telescope, and found an angular diameter of $\sim$$3''$. 
\HST\/ imaging was carried out at
optical wavelengths using the original Wide Field and Planetary Camera
(WF/PC1) and the Faint Object Camera (FOC) by Bianchi et al.\ (1995).
Adams et al.~(1984) published an 
{\it International Ultraviolet Explorer\/} (\IUE\/) spectrum,
which has been reanalyzed subsequently by others, 
including Pe\~na,
Torres-Peimbert, \&  Ruiz (1992), Bianchi et al.~(1995), and
Henry, Kwitter, \& Howard (1996).  
Heber, Dreizler, \& Werner (1993) reported on high-resolution ultraviolet
spectra of the K\,648 central star obtained 
with \HST\/ and the Goddard High Resolution Spectrograph (GHRS)\null. 
McCarthy et al.~(1997) analyzed high-resolution, high signal-to-noise optical
spectra obtained with HIRES on the Keck 10-m telescope.

The distance and reddening toward K\,648 are 
well constrained owing to its association with M15.
The foreground reddening toward M15 is $E(B-V)$ = 0.10~$\pm$~0.01~mag, as
estimated by several independent techniques (Durrell \& Harris 1993).
The reddening derived from the nebular spectrum of K\,648 agrees well
with this 
value.  The four published measurements of the Balmer 
decrement in K\,648 tabulated by Adams et al.~(1984) yield an average 
H$\alpha$/H$\beta$ = 3.23~$\pm$~0.16.  Adopting the Case~B
recombination ratios from Osterbrock (1974), and the reddening
law of Cardelli, Clayton, \& Mathis (1989), we estimate
$E(B-V) = 0.13\pm0.05$~mag (assuming an electron 
temperature of 1--$2\times10^{4}$~K)\null.
Adjusted for the Durrell \& Harris~(1993) reddening,
the {\it Hipparcos\/} subdwarf calibration of Reid (1997) yields a
true cluster distance modulus 
$(m-M)_{0} = 15.45~\pm~0.1$~mag, 
or a distance of $12.3\pm0.6$~kpc.

The chemical abundances of M15 red giants 
and K\,648 have
been the subject of numerous studies.
Estimates of the metal abundance of red giants in M15 range from
$\rm[Fe/H] = -2.4$ (Sneden et al.~1997) up to
$-2.17$ (Zinn \& West 1984) and
$-2.12$ (Carretta \& Gratton 1997).
The abundances of neon and argon
in K\,648 are subsolar by similar factors
of $\sim$100 (Barker 1980; Adams et al.~1984).  However,
K\,648 has enhanced oxygen and 
a very high carbon abundance (Howard et al.\ 1997),
implying that
some stellar core material has been ejected into the nebula. 
It is possible that the K\,648 progenitor experienced
a ``third dredge-up'' of core material following
a thermal pulse on the
AGB (e.g., Buell et al.~1997; Sweigart 1999), even though it has been
shown theoretically 
that third dredge-ups do not occur in model stars with initial
masses below $\sim$$1 \, M_{\odot}$ (Lattanzio 1989).
Hydrodynamical simulations of CE evolution show the
development of circulation patterns 
(Bodenheimer \& Taam 1984; Livio \& Soker 1988),
which could mix core material to the surface (Iben \& Livio 1993).
Thus close-binary evolution is an alternate mechanism 
for carbon enrichment in K\,648.  

\section{\leavevmode\pmb{\HST\/} Observations}

Motivated by the above considerations, we searched for periodic light 
variations of the K\,648 central star.  As noted,
such variations 
can arise in a close binary
from heating effects on a nearby main-sequence companion of the hot 
central star.
Because of the difficulty of accurate 
photometry of the K\,648 central star using ground-based techniques,
due to the very crowded field as well as the presence of a bright surrounding 
PN,
we carried out our monitoring with \HST\/ and
its
WFPC2 camera.

We observed K\,648 
during 10 \HST\/ orbits between
1998 December~15 and~22 UT\null.
We obtained a total of 79 images, with
K\,648 placed 
near the center of the PC chip.   Most of the observations (57)
were optimized for 
photometric monitoring of the central star, and were made
with the broad- and intermediate-band 
F336W, F439W, F547M, and F814W filters (which we denote $U$, $B$, 
$y$, and $I$ respectively). 
The exposure times in these filters were 23, 20, 12, and 14~s, 
respectively.
The $y$ filter
was chosen rather than $V$ since the former 
suppresses the strong [\ion{O}{3}] 5007~\AA\ nebular emission
line, and we included the $I$ filter since the amplitude of the heating 
effect on a cool companion of a hot star increases with wavelength.
We employed a geometrically increasing series
of time spacings that yielded near-optimal 
sensitivity to periodic light variations over a period range
of $\sim$45 minutes to $\sim$7 days.
Some of the exposures (7 of the 57) were, however,
affected by intense cosmic-ray hits during spacecraft
passages through the South Atlantic Anomaly,
and thus were unusable.  

The remainder of each of the \HST\/ orbits was used to obtain
observations of
K\,648 through 
the F502N, F656N, and F658N narrow-band filters, which are centered on the
[\ion{O}{3}], H$\alpha$, and [\ion{N}{2}] nebular emission lines. These were
obtained in order to 
study the morphology of the PN at WFPC2 resolution.
The multiple exposures 
obtained in these filters totaled 17.3,
190.3, and 15 minutes, respectively.

\section{Central Star Photometry}

Aperture photometry of the K\,648 central star in each of the usable
broad-band images was obtained using DAOPHOT (Stetson 1987).
The results showed no
significant variability of the nucleus of K\,648 over the 
seven-day \HST\/ observing campaign, at the $\simlt 1\%$ detectability
level of our data.  Although a positive result would have established the 
binarity of the central star, our negative result does not rule it
out.
The nucleus could still be a binary, either viewed close to pole-on, 
or of sufficiently large orbital period and separation for the heating effects 
to be negligible.  If the star is a binary, a precise lower limit
on its possible orbital period is difficult 
to state based on our null detection, 
since the limit depends both on the viewing geometry and the radius of the 
putative companion.  However, 
the lower limit is probably of order several 
days.

We then combined all of the 
usable images in each broad- and intermediate-band 
filter 
using a median algorithm,
and rederived the aperture photometry from
the resulting composite images.  Due to the strong cluster background 
in our frames, it was not necessary to apply charge-transfer-efficiency (CTE)
corrections.
Aperture corrections were determined
from several nearby bright, isolated stars in the PC images.  
The errors associated with the aperture
corrections ($\sim$1\%) dominate over the DAOPHOT-reported Poisson errors
($\sim$0.1\%) in the magnitudes from the combined images.
By a careful comparison of each broad-band
image with the H$\alpha$ image, we estimate that the magnitudes
are contaminated by nebular line emission at no more than the $\sim$1\% level 
in $U$, $B$, and $y$, and $\sim$3\% in $I$\null.   
The central star colors were then 
transformed to the standard 
Johnson-Kron-Cousins {\it UBVI\/} system, according to the
prescription of Holtzman et al.~(1995).  The calibrated magnitudes of the
K\,648 central star are summarized in Table~1.  The systematic errors of the 
calibrated magnitudes and colors are of order 2--3\% (Holtzman et 
al.~1995). The conversion from {\it UBVI\/} magnitudes to absolute
fluxes, also given in Table~1, was calculated
according to Bessell, Castelli, \& Plez (1998).

In order to study the spectral energy distribution (SED) of the central 
star, we will consider both the \HST/WFPC2 flux data and the 
ultraviolet spectrum. 
As discussed by Adams et al.\ (1984), 
flux contamination in the large-aperture \IUE\/ spectrum
from neighboring stars and the nebula is small at $\lambda < 1800$~\AA\null.
Therefore, we restricted 
our analysis to the short-wavelength spectrum. We used the highest-S/N
\IUE\/ 
observation available in the archive maintained at STScI,
image number SWP~17069.  
The mean fluxes in 100~\AA-wide passbands centered at 
$\lambda$ = 1300, 1450, 1650 and 1800~\AA\ are 
listed in Table~1.
These passbands were
selected to avoid obvious emission and absorption lines, and thus
represent the central star's continuum.

The K\,648 central star was also observed
with the \HST's Faint Object Spectrograph (FOS).  The 
unpublished \HST/FOS
spectrum was kindly provided in electronic form
by R.~Bohlin (2000), calibrated on
the white dwarf scale according to
Bohlin (1996) and Bohlin, Lindler, \& Keyes (1995).
The uncertainty in FOS absolute fluxes 
are estimated to be $\sim$5\% for
this spectrum.  In a direct comparison
at short wavelengths, the
\IUE\/ spectrum appears to suffer from a $\sim$15\% loss of
flux relative to the FOS data, 
possibly due to misalignment in the slit.  The mean
\IUE\/ and \HST/FOS
fluxes at $\lambda$ = 1300, 1450, 1650 and 1800~\AA\ are
compared in Table~1.  Given the flux discrepancy and the superior
resolution of \HST/FOS,
we employ only the FOS data in our
analysis of the SED below.  The average continuum flux
data points are sufficient to constrain the various 
physical parameters that we seek from
the central star's SED.

We adopt the reddening law of Cardelli, Clayton, \& Mathis (1989)
to deredden the \HST/WFPC2 and FOS flux data.
The ratios of total extinction ($A_{\lambda}$)
to the total visual extinction ($A_{V}$) 
for each of our flux data points
are listed in Table~1. 
As noted 
above, the reddening for M15 is known to be
$E(B-V) = 0.10\pm0.01$~mag from Durrell \& Harris~(1993).
Having assumed $R_{V} = A_{V}/E(B-V) = 3.1$, we previously estimated
the cluster's distance using a
visual extinction of $A_{V} = 0.31$ mag.
In our analysis below, we will consider possible variations in
the extinction law ($R_{V}$).  Thus,
for simplicity, we will hold $A_{V} = 0.31$ mag constant
so that the cluster is always assumed to be 
at the same distance.  Plausible variations in $R_{V}$ while
$A_{V}$ is held constant
lead to only small changes in
$E(B-V)$, but relatively large changes in the
dereddened far-ultraviolet fluxes.
In our analysis below, we will
assume that the K\,648 central star 
has the same reddening as the cluster, but consider also
the possibility that the central star is additionally reddened
(i.e., perhaps by its own ejected material).

\section{Luminosity, Effective Temperature, and Mass of the K\,648 Central
Star} 

With our new \HST/WFPC2 and FOS data and the recent 
{\it Hipparcos}-calibrated distance to M15, it is worth
revisiting the luminosity
and color-temperature calculation for the K\,648 central star given by
Adams et al.\ (1984). Previous ground-based photometry of the nucleus required 
large and uncertain corrections for nebular contamination, which are of course 
unnecessary with the WFPC2 data.
Figure~1 shows the SED of the central star. We plot 
both the new {\it UBVI\/} points from our WFPC2 observations 
and the four points from the 
short-wavelength FOS data. 

Our approach is to fit Lejeune et al.~(1997) model stellar
spectra\footnote{We use their ``corrected'' model spectra
and adopt $\log g$ = 5.0, for which model spectra are
provided for the entire range of effective temperatures of interest.
The specific choice of $\log g$ has a negligible effect
on our derived results.
In order to convert to astronomical units, we assume values
for the Sun of $\log L_{\odot} = 33.582$ and
$\log R_{\odot} = 10.842$, where $L_{\odot}$ and $R_{\odot}$
are in units of $\rm erg \, s^{-1}$ and cm, respectively.}
to the flux data summarized in Table~1.
Each model spectrum is tabulated as wavelength ($\lambda$)
in nm and flux moment
($H_{\nu}$) in $\rm erg \, s^{-1} \, cm^{-2} \, Hz^{-1}$,
which we convert to flux using the formula
\begin{equation}
f_{\lambda} = \frac{ 0.4 c H_{\nu} }{\lambda^2 },
\end{equation}
where $c$ is the speed of
light in $\rm nm \, s^{-1}$, and the units of $f_{\lambda}$
are $\rm erg \, s^{-1} \, cm^{-2} \, \AA^{-1}$\null.  At Earth the
flux from the star is given by
\begin{equation}
F_{\lambda} = \left( \frac{ \pi R^2 }{ D^2 } \right) f_{\lambda},
\end{equation}
where $R$ is the radius of the star and $D$ is its distance.
In our case, $D$ is known, $F_{\lambda}$ has been measured (Table~1), and 
$f_{\lambda}$ can be obtained as a function of $\Teff$ from the 
Lejeune et~al.\ model spectra as given above. Thus, with 
multi-wavelength data, we can 
solve simultaneously for the stellar radius and effective temperature.
We adopt the model fluxes
at the effective wavelengths of the various bandpasses.
The luminosity of
the central star ($L$ in units of $\rm erg \, s^{-1}$) is then given by
\begin{equation}
L = 4 \pi R^2 \sigma T_{\rm eff}^4
  = 4 \pi^2 R^2 \int_{0}^{\infty} f_{\lambda} d\lambda.
\end{equation}

In our first round of fitting, a $\chi^2$
minimization is used to find $R$, $\Teff$, and the extinction ratio
$R_{V}$.  We employ model spectra with
$\rm[Fe/H] = -2.0$, which corresponds roughly to the metallicity
of M15 red giants.
The very low metallicity assumed here may not be appropriate for the central
star if its photosphere contains material dredged up from the core, so
we also fit using model spectra with $\rm[Fe/H] = 0.0$, which however yielded
the same results.  We conclude
that the color temperature is insensitive to different assumptions
for the K\,648 central star photospheric abundances.
In this round of fitting, the 1$\sigma$ confidence intervals are
$T_{\rm eff} = 37$,500 to 50,000~K, and
$\log(L/L_{\odot}) = 3.71$ to 4.11.  However, the upper limits are
set by the highest temperature model spectrum fitted and not the
$\chi^2$ statistic.  The best-fit for $R_{V}$ is quite low,
$R_{V} = 2.0_{-0.2}^{+0.4}$ (1$\sigma$ errors), 
but within the range of known
values ($R_{V} = 2$ to 5; Fitzpatrick 1999).

We considered the possibility that the K\,648 central star is reddened in
excess of the mean cluster reddening, perhaps due to dust in the surrounding
nebula. This possibility is raised by our measured $B-V$ color of $-0.13$.
For example, the Lejeune et al.\ intrinsic color of $-0.31$ at 40,000~K then 
suggests $E(B-V)\simeq0.18$~mag. We also recall that the nebular Balmer
decrement, as discussed in \S2, indicates $E(B-V) = 0.13 \pm 0.05$~mag.
However, the central star's $U-B$ and $V-I$ colors and the nebular value are
all consistent within their errors with the adopted $E(B-V) = 0.10$~mag.
In any case, we note that 
the effect of increasing the K\,648 central star
reddening will be to {\it increase\/} the derived stellar luminosity and
mass, which will only strengthen our conclusions about its evolutionary
history (see \S7). 

In a second round of fitting, a $\chi^2$
minimization is used to find $R$, $\Teff$, and the 
reddening $E(B-V)$, which is allowed to be larger than the
cluster's foreground reddening of $E(B-V)$ = 0.10.  In this case,
we assume the ``standard'' extinction ratio of $R_{V}$ = 3.1.
A $\chi^2$ minimization yields the 1$\sigma$ confidence intervals of
$T_{\rm eff} = 33$,000 to 47,500~K
and $\log(L/L_{\odot}) = 3.61$ to 4.02 dex.  The best-fit
reddening is $E(B-V) = 0.13_{-0.01}^{+0.02}$ (1$\sigma$ errors).
This is in excellent agreement with that found using the nebular Balmer
decrement, as noted above.

Not surprisingly, our analysis of the SED
only weakly constrains the effective temperature 
of the central star (because even 
the far-UV points are essentially on the Rayleigh-Jeans tail).  
The stellar effective temperature is almost entirely
unconstrained by our new optical flux data, except that the lack of a 
perceptible Balmer jump weakly favors a high temperature. 
The far-UV data are sensitive to the
adopted reddening and reddening law, in addition
to the effective temperature
of the central star.

Fortunately, spectroscopic analyses 
based on absorption-line measurements can provide 
much tighter constraints. Analyses of this type
have yielded $T_{\rm eff} = 37,000$~K
(Heber et al.~1993) and $T_{\rm eff} = 43,000 \pm 2,000$~K 
(McCarthy et al.~1997; M\'endez 1999), based on UV and optical spectra,
respectively.
These spectral-line temperatures are
consistent with our color temperature, but more accurate, and we
give them high weight. They are also consistent with the value of 38,000 
$\pm$ 4,000~K adopted by Adams et al.\ (1984) on the basis of the Zanstra and 
Stoy nebular methods and their analysis of the 
optical/UV energy distribution. 
We therefore
adopt an effective temperature of
$T_{\rm eff} = 40,000 \pm 3,000$~K\null. 

With the temperature fixed to $T_{\rm eff} = 40,000$~K, 
and allowing $R_{V}$ to vary, we find a best-fit of
$R_{V}$ = 2.3.  For $R_{V}$ = 3.1 and allowing $E(B-V)$ to vary 
while 
the temperature is held fixed, we find a best-fit
of $E(B-V) = 0.13$.  
Given the low value of $R_{V}$ = 2.3 found, the scenario in
which the K\,648 central star is reddened slightly more
than the mean reddening of the cluster
is probably preferred.
However, both scenarios yield consistent
results for the luminosity and radius of the
K\,648 central star.
Our adopted effective temperature of
$T_{\rm eff} = 40,000 \pm 3,000$~K corresponds
$\log(L/L_{\odot}) = 3.78 \pm 0.08$,
and $R/R_{\odot} = 1.61_{-0.06}^{+0.11}$.
Figure 1 shows the SED for
a standard reddening law of $R_{V}$ = 3.1 and a
reddening of $E(B-V) = 0.13$.

An estimate of the K\,648 central star mass
follows from theoretical calculations of 
post-AGB evolutionary tracks
(Pacz\'{y}nski 1971; Sch\"{o}nberner 1979, 1983; Vassiliadis \& Wood 1994).
The well-known luminosity/core-mass relation is one of the basic
predictions in AGB and post-AGB stellar evolution theory.
Vassiliadis and Wood find a luminosity/core-mass relation of
$L/L_{\odot} = 56694(M_c/M_{\odot} - 0.5)$, 
independent of metallicity, and applicable to both hydrogen- and 
helium-burning tracks.  This relation predicts the luminosities of their 
individual post-AGB tracks with an rms 
scatter of $\pm0.02$~dex in $\log(L/L_{\odot})$.
Using this formula, we find that 
the mass of the K\,648 central star is
$M = 0.60 \pm 0.02 \, M_{\odot}$,
where the uncertainty
does not include the systematic error
associated with the theoretical luminosity/core-mass
calibration (which is comparable to the 
statistical error).  

For our derived mass and radius, 
the surface gravity of the central star is predicted to be
$\log g = 3.8$. This agrees 
well with both
Heber et al.~(1993) and McCarthy et al.~(1997), who found $\log g$'s of 4.0
and 3.9, respectively,
from their spectroscopic analyses.

\section{Morphology of the Nebula}
 
K\,648 is barely resolved optically from the ground (see, for
example, G\'orny et~al.\ 1999).  \HST\/ imaging with the pre-repair WF/PC1 was
reported by Bianchi et~al.\ (1995).  Our new WFPC2 data confirm the basic
morphological features discussed by Bianchi et~al., but also provide new
information revealed by the even higher spatial resolution of the repaired
\HST.
 
K\,648 shows three main structural features, which
are revealed clearly in our new images and
can be described approximately as follows: (1)~a very bright, hollow, inner
elliptical shell, with outer major and minor axes of $0\farcs8$ and $0\farcs6$;
(2)~a larger, somewhat fainter, and fairly sharp-edged elliptical
shell of dimensions $3\farcs1 \times 2\farcs7$; and (3)~a much fainter,
diffuse elliptical halo that fades away gradually with radius but can be
detected out to dimensions of at least $6\farcs5 \times 5\farcs5$.  The
corresponding dimensions in parsecs are: inner shell $0.05 \times 0.04$~pc,
outer shell $0.19 \times 0.16$~pc, and faint halo $0.38 \times 0.33$~pc.
These
three features are illustrated in our deep WFPC2 H$\alpha$ image of K\,648
displayed in Figure~2, where the image stretch is logarithmic and has been
adjusted to show the faint outer halo and (overexposed in this stretch) the
outer sharp-edged elliptical shell. The structure of the bright inner shell is
represented in Figure~2 by superposed isophotes. The two bright inner
elliptical shells seen in K\,648 are a typical morphological feature of many
``elliptical'' PNe (e.g., Balick 1987). Large faint halos are
also often seen when sufficiently deep images are available.

The most remarkable feature of the morphology, seen most clearly
for the first time in our WFPC2 images, is a bright arc on the northwestern
end of the nebula's major axis,
located just inside the edge of the outer bright elliptical
shell. This arc is especially prominent in the light of [\ion{N}{2}], and is
seen as well in H$\alpha$ and [\ion{O}{3}]\null. A corresponding feature at the
other end of the major axis does not appear to be present, although two fairly
bright stars are unfortunately superposed at exactly that position. Figure~3
is a false-color image that illustrates this feature.
 
At first sight, one might interpret the arc as a bow
shock, arising from the high velocity of M15 through the interstellar medium
(ISM)\null.
The PN in the globular cluster M22 shows
strong evidence for an interaction
with the ISM in ground-based images
(Borkowski, Tsvetanov, \& Harrington 1993), and an ISM interaction is also
suspected involving
the PN in NGC~6441 based on \HST\/ WFPC2 images (Jacoby et al.\ 1997).
However, we believe that an interaction is an
unlikely explanation for the arc in K\,648,
because the faint outer halo of the PN appears completely
undisturbed.  Knowledge of the absolute proper motion of M15 would also be
useful in assessing this possibility, but recent ground-based studies have not
even agreed on the signs of the right-ascension and declination components
(Scholz et al.\ 1996 and references therein). However, three out of the four
recent studies summarized by Scholz et~al.\ have found a southward declination
component. In this case, the arc-like feature is on the wrong side of K\,648
for it to be a bow shock.  Actually it would be
surprising to find evidence for an ISM
interaction in K~648, since M15 lies $\sim$5600~pc below the Galactic plane,
compared with 400~pc for M22 and 1000~pc for NGC~6441.
 
The arc
feature is probably composed of ``FLIERs'' (fast low-ionization
emission regions), which are a phenomenon
seen in many double-shell elliptical PNe. FLIERs
often occur at or just inside the outer elliptical
shell, and often near the ends of the
major axis, just as seen in K~648.
The origin of FLIERs is still uncertain (see Balick et al.\ 1998
and references therein).  It is somewhat unusual to find FLIERs at only one
end of the major axis, as in K\,648.
However, M~2-2 is one example of a field PN that
exhibits a concentration of FLIERs on only one side of the
nebula (Manchado et al.~1996, pp.~107 and 147).
M~2-2 even shows a faint ``neck'' connecting its inner and outer
elliptical shells, along the major axis in the direction of the FLIERs, which
is seen also in our Figure~3.

In summary, the morphology of K\,648 is fairly typical of that seen in many 
elliptical PNe, with the only somewhat unusual feature being a concentration 
of FLIERs at one end of its major axis.  There is no strong evidence for 
interaction with the interstellar (or intracluster) medium.  Whether the
morphologies of all elliptical PNe are due to binary-star interactions has
been debated vigorously (e.g., Kastner, Soker, \& Rappaport, eds., 2000), but
the answer remains uncertain.  However, nebular ejection as a 
consequence of a CE interaction, or the enhanced rotation of the AGB 
progenitor that would result from tidal spinup in a binary system or from a 
stellar merger, are natural ways to produce the density contrast seen in 
elliptical PNe. Thus the morphology of K\,648 may suggest that it is descended 
from a close-binary progenitor.

\section{The Anomalous Mass of the K\,648 Central Star} 

The mass we have derived for the K\,648 central star,
$M = 0.60 \pm 0.02 \, M_{\odot}$, is anomalously
high for a stellar remnant in an old globular cluster (or in the Galactic 
halo), as
shown by several recent {\it HST}/WFPC2 studies.
Richer et al.~(1997) observed a large number of white dwarfs
in the GC M4.  These authors found the M4
distance modulus by fitting subdwarfs to the cluster
main sequence, and then derived a model-dependent mean white dwarf
mass of $M_{\rm WD} = 0.51 \pm 0.03 \, M_{\odot}$.   Cool et al.~(1996)
observed a smaller number of white dwarfs in NGC~6397.  They
also derived the cluster distance modulus by main-sequence fitting, and
estimated a model-dependent
white dwarf mass of $M_{\rm WD} = 0.55 \pm 0.05 \, M_{\odot}$.
Renzini et al.~(1996) 
observed the white dwarf sequence in the globular
cluster NGC~6572.  However, instead of measuring the white dwarf masses
based on the cluster distance,
these
authors adopted 5 field halo white dwarfs with measured trigonometric 
parallaxes
as calibrators.  The halo white dwarfs have a mean
$M_{\rm WD} = 0.51 \pm 0.01 \, M_{\odot}$, based on fitting to theoretical 
cooling curves.  From this calibration, Renzini et al.\
derived a distance modulus of 
$(m-M)_0 = 13.05 \pm 0.10$ mag by fitting
to the cluster white-dwarf sequence.  
The Renzini et al.~(1996) distance modulus
compares
with $(m-M)_0 = 13.23 \pm 0.10$ mag found by Reid (1997)
who fit subdwarfs to the
main sequence (both authors adopted the same reddening).
We can recalculate
the masses of white dwarfs in NGC~6572 relative to the main sequence
as follows.  Referring to the same model white dwarf sequences
(see Bergeron, Ruiz, \& Leggett 1997) employed
by Cool et al.~(1996) and Richer et al.~(1997), we find that at a fixed color,
the white dwarf mass scales with absolute magnitude according to
$M_{V} \propto 2.75 M_{WD}$.
Allowing for the Reid (1997) distance,
the absolute magnitude of the cluster white dwarfs shift
by $\Delta M_{V} = -0.18$ mag, which yields a shift
of $\Delta M_{\rm WD} = -0.06 M_{\odot}$ relative to the
local calibrators employed by Renzini et al.~(1996).
Thus, we find $M_{\rm WD} = 0.46 \pm 0.04 M_{\odot}$
for the mass of white dwarfs in NGC~6572, where the associated error
is dominated by the uncertainty in the distance modulus.

Taking the weighted
average of these three measurements in GCs, we derive a best estimate
for a mean globular cluster white-dwarf mass of 
$M_{\rm WD} = 0.50 \pm 0.02 M_{\odot}$.
This mass is consistent with the mean mass of the halo field white
dwarfs selected by Renzini et al.~(1996) as calibrators.  Thus we conclude 
that the masses of white
dwarfs in GCs and in the halo
are significantly lower than the mass we estimate for the
K\,648 central star.  

Yet another argument
for a high mass for the K\,648 nucleus
arises from a comparison of nebular and stellar
evolutionary timescales (see, e.g., Bianchi et al.~1995; Buell et al.~1997).
If we assume a nebular angular
radius
of $r\simeq1\farcs5$ (see \S6; this radius refers to the
nebula's outer
bright elliptical shell, which we interpret as having arisen at
the main ejection event whose onset occurred at the
end of the AGB phase),
and adopt
a typical PN expansion rate of $v_{\rm exp}\simeq20\kms$,
then the dynamical age of K\,648 is
\begin{equation}
\tau_{\rm dyn} \ \simeq \ 4400 \, {\rm yr} \ \left(\frac{r}{1\farcs5}\right)
\ \left(\frac{v_{\rm exp}}{20\kms}\right)^{-1}.
\end{equation}
To our knowledge, post-AGB evolutionary models do not exist for
the low metallicities and low progenitor masses relevant for K\,648.
However, indications are that the nucleus can evolve from the AGB
to a temperature of $\sim$40,000~K 
within $\sim$4400~yr only if its mass exceeds $\approx0.59M_\odot$ 
(see, e.g., Sch\"onberner 1983;
Vassiliadis \& Wood 1994; Bl\"ocker \& Sch\"onberner 1997; and references 
therein).  A slightly lower mass near $0.56 M_\odot$
is possible for a helium-burning central star, according to the Vassiliadis \& 
Wood calculations.
However, at the considerably
lower core masses implied by the \HST\/ observations of 
white dwarfs in GCs, 
the evolutionary timescales are likely to be extremely 
leisurely and greatly exceed $\tau_{\rm dyn}$.

The weight of the evidence thus suggests strongly 
that the mass of the K\,648 central star 
is unusually high compared to the normal
remnants of halo and GC stars.  
Interestingly, the central star of the PN in 
M22 also appears to be much
more 
luminous than would be expected if it is the remnant of a turnoff star of 
$\sim$$0.8M_\odot$, according to an analysis of ground-based and \IUE\/ spectra
by
Harrington \& Paltoglou (1993). However, their determination of the star's 
high $L/L_\odot$ depends on
the uncertain $\Teff$ of 
the star. Moreover, this object presents additional evolutionary puzzles 
because of the extremely
hydrogen-deficient composition of its nebula but not of the 
central star itself, the latter being 
a helium-rich sdO star.\footnote{For completeness, we 
note that the luminosities of the central stars of
the PNe in Pal~6 and NGC~6441 suggest that they may not be as massive as the
nuclei of K~648 or the PN in M22. 
However, the luminosities of these central stars are
not known with great accuracy, so the constraints on their
masses are not tight (see discussion in Jacoby et al.\ 1997). Improved
observations of these two objects, along with stronger evidence that the PN
associated with Pal~6 is actually a cluster member, would be very useful.} 

\subsection{Mass Augmentation in a Close Binary?}

We believe
the most likely explanation for the unusually high mass of the K\,648 
nucleus is that the 
star has experienced mass augmentation at some point in its evolutionary 
history, which 
would be explained most naturally through an interaction in a close binary 
system.

In M15 the current main-sequence turnoff mass is $\sim$$0.80~M_{\odot}$ (Reid
1997). The initial mass of the K\,648 progenitor that would be implied by a
current mass of $0.60 \, M_\odot$ is quite uncertain, because the theoretical
initial-final mass relation depends on metallicity, and we require an
extrapolation to lower metallicities than provided by the available post-AGB
models.  However, by inspection of the lowest-metallicity post-AGB tracks of
Vassiliadis \& Wood (1994), we conclude that the most probable initial mass
of the K\,648 progenitor is $M_{i} \simlt 1.0~M_{\odot}$.  In this case, the
original $0.80~M_{\odot}$ star needs to have experienced a modest $\sim$25\%
mass augmentation, and then evolved as a higher-mass star to become K\,648.
This higher mass may also be consistent with the high carbon abundances in
K\,648, which, as noted above, may have resulted from the dredge-up of core
material due to the thermal-pulse instability on the AGB (Lattanzio 1989) of
a star of $\sim$$1\,M_{\odot}$. 

If the PNe in GCs arise from close binaries that merged on the main sequence,
then they are the descendants of blue stragglers.  This may lead to a problem
in the predicted numbers of PNe in GCs, as shown by the following rough
estimate. If we assume that the total number of blue stragglers in GCs is of
order 100 per cluster, or about $10^4$ in total, that each has a lifetime of
$\sim$$10^9$ yr, and that each PN lives $\sim$$2.5\times 10^4$ yr, we predict
there should be only $\sim$$0.25$ PNe in the GC system.  This suggests that
the blue stragglers (i.e., the Case~A mergers)
may not be able to account for all four PNe known in GCs.
Scenarios in which the PN is ejected during 
a Case~B or~C CE interaction, leaving 
a nucleus that is still a binary (e.g., Han et al.\ 1995),
suffer from a failure to predict the
observed high luminosity of the K\,648 nucleus. This channel could, however,
account for about two
fainter PN nuclei in the GC system, perhaps not unlike those in 
NGC~6441 and Pal~6.
Han et al.\ also
predict that $\sim$7\% of the stars in an old population will merge in CEs, 
leaving red giants of higher masses; 
these merger products could then evolve later to produce PN nuclei
as massive as K\,648.  This
formation channel could produce about one PN in the entire GC
system.  It thus apppears that close-binary evolution may be able to account 
for the observed small number of PNe in GCs, and that merger scenarios are 
most likely to produce a high-mass central star like that of K\,648.

\subsection{Stochastic Scatter in the Initial-Mass Final-Mass Relation?}

The mass of the K\,648 central star is anomalously high compared to the mean
remnant mass in GCs. In the previous subsection we explored scenarios
involving binary interactions. Binary evolution allows us to retain the
assumption of a unique, single-valued relationship between initial and final
stellar masses for the evolution of single stars.  In this subsection we
explore whether the high mass of K\,648 instead 
simply indicates a large scatter in
the initial-final mass relationship. 

If the discrepancy between the mass of the K\,648 nucleus and the mean mass 
of white dwarfs in GCs were due to stochastic mass loss, then the standard 
deviation in the final masses would be of order $0.1 \, M_\odot$. 
The best-observed white-dwarf sequence in a GC is that of Richer et al.\ 
(1997), whose data do not exclude a scatter this large given the 
photometric errors.

Reid (1996) has estimated the initial and final masses
of white dwarfs in several open clusters, and argued for appreciable 
stochastic scatter in the relationship.  For
the Hyades and Praesepe clusters with initial masses
of $M_i \sim 2.5 M_{\odot}$, he finds 
a mean final mass of $M_f = 0.69 \, M_{\odot}$, with a standard
deviation of $0.13\,M_{\odot}$.  

Thus we cannot rule out the possibility
that the anomalously high mass of the K\,648 nucleus 
is due to stochastic mass loss.  However, as discussed above, binary-star 
processes appear to be a sufficient explanation.
In fact,
the scatter in final masses reported at higher progenitor
masses could in itself also arise from close-binary interactions, 
as we have suggested for
the case of K\,648. 

\section{Conclusion}

K\,648 in the GC M15
is the prototype of PNe formed by low-mass and low-metallicity
progenitors. 
We have presented new observations of K\,648 obtained with \HST\/ and the
WFPC2 camera, from which (combined with our analysis 
and discussion of previously published information)
we have
calculated the central star's 
temperature, luminosity, and mass.
We have also presented the morphology of
K\,648 with unprecedented resolution.  It is a fairly typical 
double-shelled elliptical PN, surrounded by a 
fainter smooth outer halo.  Its only 
unusual structural feature is 
a bright arc at one end of the major axis, which is bright in [\ion{N}{2}] and 
is probably 
a collection of FLIERs such as seen in other elliptical PNe.  

The derived mass of $0.60 \, M_\odot$ 
is consistent with that required for 
the remnant to be able to
evolve to high enough temperature to ionize its surrounding 
envelope before the envelope has had time to dissipate.  
However, this mass is 
significantly
higher than those of white dwarfs in GCs.
We outlined several possible evolutionary 
scenarios through which mass 
transfer or mergers may lead to a luminous PN central star in a GC.

In some of these scenarios, the central star would still be a close binary.
However, our WFPC2 observations revealed
no significant photometric variability
over periods of 45 minutes to 7 days.  This null result does not
rule out the possibility that the nucleus is a binary, either seen close to 
pole-on, or having 
an orbital period of more than several days.  However, we suggest that 
the present central star is probably descended from a binary that 
merged on the main sequence or in a common-envelope interaction.

\acknowledgements

Support for this work was provided by NASA through grant number
GO-06751.01-95A from the Space Telescope Science Institute, which is operated
by the Association of Universities for Research in Astronomy, Inc., under NASA
contract NAS5-26555. Brian Warner provided support for this program at a
critical time. Laura Fullton and Karen Schaefer assisted in the initial
discussion and planning of the observations, and Denise Taylor dealt
cheerfully with implementation of the complex and demanding \HST\/ scheduling
requirements. Roberto M\'endez and Ulrich Heber sent useful private
communications.   We thank Ralph Bohlin for providing his unpublished,
calibrated \HST/FOS spectrum of the K\,648 central star.

\clearpage

\begin{figure}
\plotone{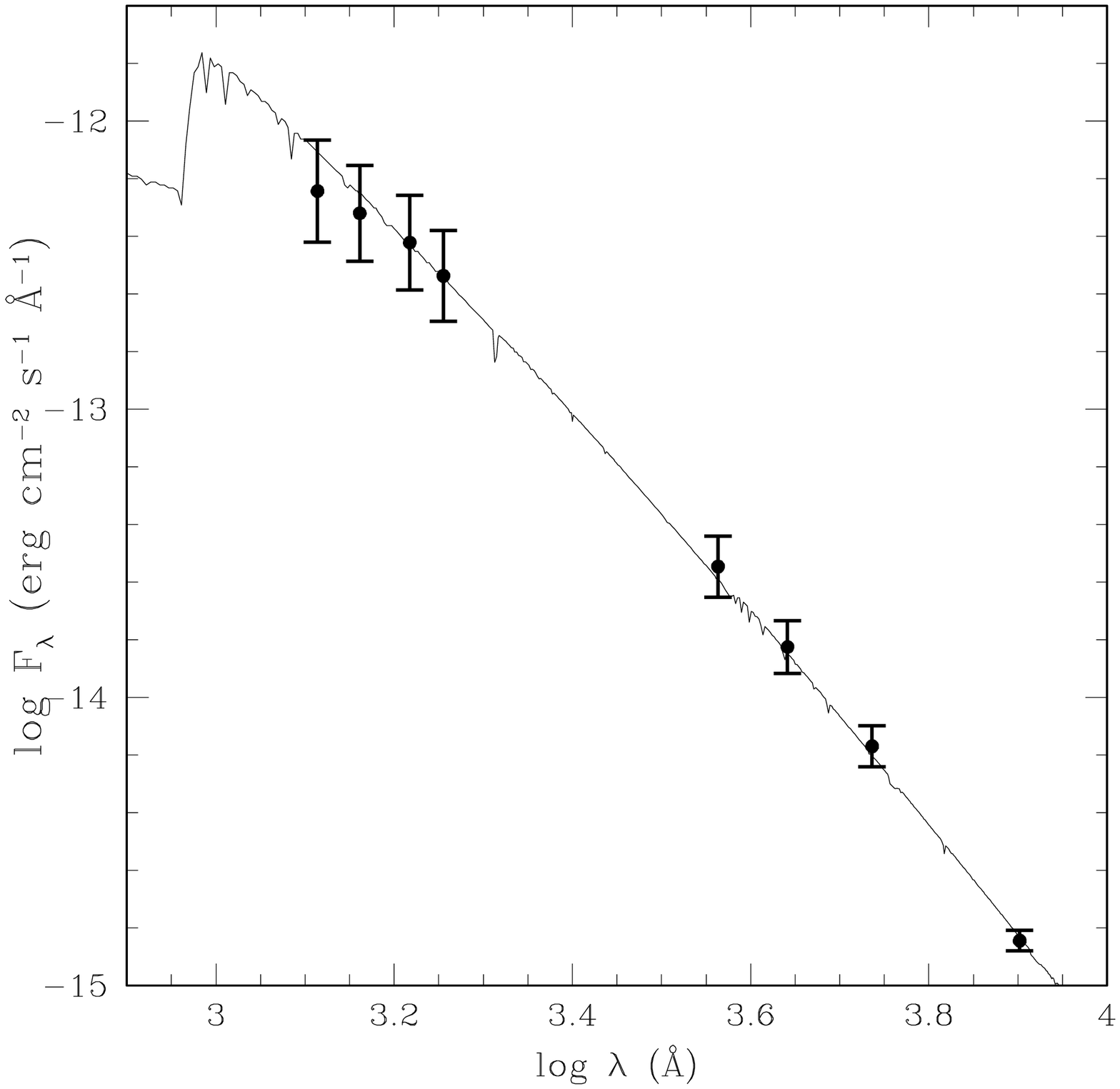}
\caption{Dereddened \HST/WFPC2 and FOS flux data points 
for the central star of K\,648 are shown with
error bars corresponding to $\pm 0.01$ in $E(B-V)$.
The error bars illustrate the relative sensitivities of 
the \HST/WFPC2 and FOS data to uncertainties in 
the reddening and the respective weights 
employed in our SED fitting trials (the optical flux data
are given the highest weight).  We adopted a
foreground reddening of $E(B-V)$ = 0.10, and found a
best-fit total reddening of $E(B-V)$ = 0.13; the excess
presumably 
arises from the K\,648 nebular material.
Overplotted is a model spectrum from Lejeune 
et al.~(1997) for $\rm [Fe/H] = -2.0$, $\log g = 5$, and 
$\Teff = 40,000$~K{}.  The
implied luminosity of the K\,648
central star is $\log(L/L_{\odot}$) = 3.78 $\pm$ 0.08.}
\end{figure}


\begin{figure}
\vskip-1in
\plotone{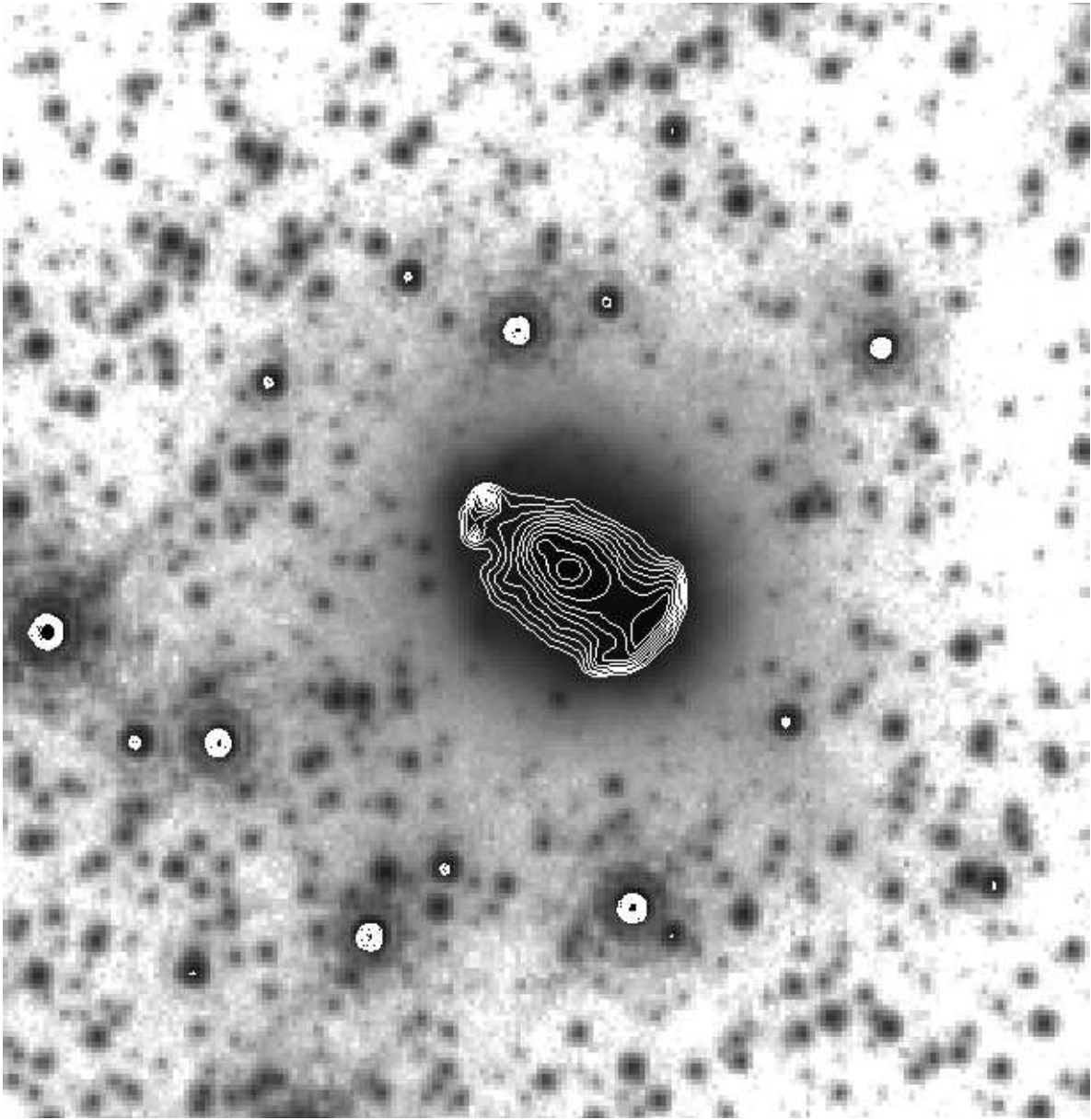}
\vskip0.3in
\caption{Deep WFPC2 image of K\,648 in the light of
H$\alpha$.  A logarithmic stretch has been chosen to show the very faint outer 
halo around K\,648, and the overexposed outermost of the nebula's bright
elliptical shells.  The inner elliptical shell is shown by superposed 
isophotes; here the contours are logarithmic,
in 10 steps of 0.1 dex.  The field of view is
210$\times$210 pixels, or 8.4 arcsec on a side.
North is 11 degrees clockwise from the right, and east is 11 degrees clockwise 
from straight up.
}
\end{figure}


\begin{figure}
\vskip-0.5in
\hskip2in
\plotone{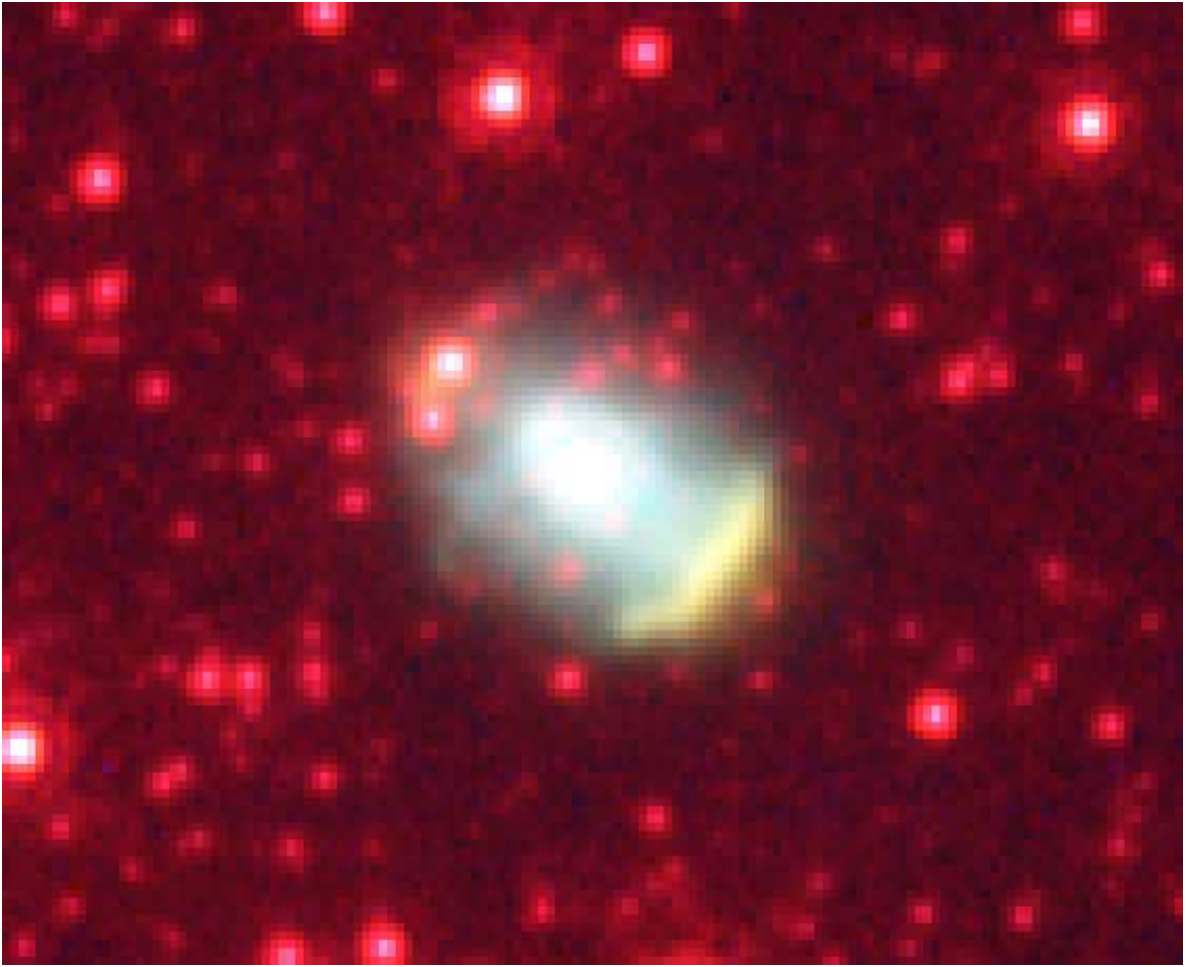}
\vskip0.3in
\caption{False-color WFPC2 image of K\,648 in the light of [\ion{O}{3}] 
5007~\AA\
(blue), H$\alpha$ (green), and [\ion{N}{2}] 6583~\AA\ (red).  
Intensity is log-scaled, and the
scale and orientation can be inferred from Fig.~2. This figure illustrates the 
inner and outer elliptical shells and the
 bright arc of ``FLIERs'' on the northwestern edge of the outer elliptical 
shell.  In this representation the large outer halo is too faint to be seen.} 
\end{figure}

\clearpage

\begin{deluxetable}{ccccccc}
\tablewidth16cm
\tablecaption{K\,648 Central Star Flux Measurements}
\tablehead{
\colhead{$\lambda$ (\AA)} &
\colhead{Mag.\tablenotemark{\ a} } &
\colhead{$F_{\lambda}$\tablenotemark{\ b} } &
\colhead{$F_{\lambda}$ IUE \tablenotemark{\ b} } &
\colhead{$F_{\lambda}$ FOS\/ \tablenotemark{\ b} } &
\colhead{$A_{\lambda}/A_{V}$\tablenotemark{\ c}} &
\colhead{$F_{\lambda,0}$\tablenotemark{\ b,c}} 
}
\startdata
 1300 & \nodata & \nodata & $1.60\times10^{-13}$ & $1.95\times10^{-13}$ & 2.903 & $5.71\times10^{-13}$ \nl
 1450 & \nodata & \nodata & $1.56\times10^{-13}$ & $1.76\times10^{-13}$ & 2.703 & $4.79\times10^{-13}$ \nl
 1650 & \nodata & \nodata & $1.27\times10^{-13}$ & $1.42\times10^{-13}$ & 2.645 & $3.79\times10^{-13}$ \nl
 1800 & \nodata & \nodata & $1.10\times10^{-13}$ & $1.14\times10^{-13}$ & 2.520 & $2.90\times10^{-13}$ \nl
 3660 & 13.55  &  $1.59\times10^{-14}$ & \nodata & \nodata & 1.569 &  $2.84\times10^{-14}$ \nl
 4380 & 14.60  &  $9.15\times10^{-15}$ & \nodata & \nodata & 1.322 &  $1.50\times10^{-14}$ \nl
 5450 & 14.73  &  $4.66\times10^{-15}$ & \nodata & \nodata & 1.000 &  $6.76\times10^{-15}$ \nl
 7980 & 14.93  &  $1.20\times10^{-16}$ & \nodata & \nodata & 0.479 &  $1.43\times10^{-15}$ \nl
\enddata
\tablenotetext{a}{\ Magnitudes are {\it UBVI\/} on the Johnson-Kron-Cousins 
standard system.}
\tablenotetext{b}{\ Observed and dereddened fluxes 
in units of $\rm erg \, s^{-1} \, cm^{-2} \, \AA^{-1}$. }
\tablenotetext{c}{\ $A_{\lambda}/A_{V}$
from Cardelli, Clayton, \& Mathis (1989); assumes $A_{V}$/$E(B-V)$ = 3.1; dereddened
fluxes correspond to $E(B-V) = 0.13$.}
\end{deluxetable}

\end{document}